\documentclass[%
reprint,
superscriptaddress,
amsmath,amssymb,
aps,
]{revtex4-2}

\usepackage{graphicx}
\usepackage{color}
\usepackage[normalem]{ulem}
\usepackage{dcolumn}
\usepackage{bm}

\begin{document}
	\preprint{APS/123-QED}

	\title{{\Large \bf{Quantum Radiometric Calibration}}} 
	\author{Leif Albers}
	\author{Jan-Malte Michaelsen}
	\author{Roman Schnabel}
	\email{roman.schnabel@uni-hamburg.de}
	\affiliation{%
		Institut f\"ur Quantenphysik und Zentrum f\"ur Optische Quantentechnologien\\
		Universit\"at Hamburg, Luruper Chaussee 149, 22761 Hamburg, Germany}

\date{\today}

\begin{abstract} 
Optical quantum computing, as well as sensing technology based on quantum correlations are in preparation. These require photodiodes with close to perfect quantum efficiency for the detection of about $10^{16}$ {to $10^{18}$} photons per second. Already the radiometric calibration is a challenge. Here, we provide the proof of concept of the quantum radiometric calibration (QRC) method, which is based on Heisenberg's uncertainty principle, derive the formula for its precision, and exclude the potential sources of systematic error. 
As an example, we calibrate a pair of the most efficient commercial photodiode at 1550\,nm to the previously unknown system detection efficiency of ($97.20 \pm 0.37$)\,\% using 10-dB-squeezed vacuum states. This value is unexpectedly low, too low for planned gravitational wave detectors and for continuous-variable optical quantum computing. With further improved precision, our work opens up the second true end-user application of the advantage provided by quantum correlations. 
\end{abstract}
\maketitle

\noindent
{\bf 1. INTRODUCTION}\\[2mm]
Technologies for improving functionality through quantum correlations are currently receiving significant public funding. There is already one example of an end-user application: squeezed light improving the sensitivity of gravitational-wave observations \cite{Schnabel2010,Vahlbruch2010,LSC2011,Khalaidovski2012,Grote2013,GWo3a2021}. 
The absolute calibration of photodiode detection efficiency could become a second successful example.
Photodiodes are operated in the unity-gain regime. The ideal case of perfect detection efficiency ($\eta_{\rm DE}\!=\!1$) means that each photon is converted into exactly one photoelectron. Also required is negligible dark current, which we take into account through the more strictly defined `quantum efficiency' ($\eta_{\rm QE}\!\le\!\eta_{\rm DE}\!\le\!1$). With close to perfect quantum efficiency, the photon statistics of light can be mapped one-to-one into photoelectron statistics for high photon fluxes in the range of 10\textsuperscript{16}\,–\,10\textsuperscript{18}\!/s (without single-photon resolution). 
Two such photodiodes together with a balanced beam splitter and a local oscillator beam (LO) of the aforementioned photon flux that is overlapped with the signal beam form a so-called balanced homodyne detector (BHD), which is arguably the most important photoelectric detector in optical quantum technologies. A BHD is able to measure the information of an arbitrary quantum state effectively and completely (`tomographically'). This includes the zero-point fluctuations of the ground state of an optical mode, which contains {\it zero} photons. 
BHDs have been used to realise quantum teleportation \cite{Furusawa1998,Bowen2003} and Schr\"odinger cat states \cite{Ourjoumtsev2006,Neergaard-Nielsen2006}. And they are to be used to read out the optical quantum computer  \cite{Larsen2019,Fukui2022,Konno2024,Larsen2025}.
For this, photodiodes with a quantum efficiency close to one (\textgreater\,99\,\%) are required to achieve the necessary fault tolerance in the future \cite{Larsen2025}. To determine the absolute efficiency of photodiodes, national metrology institutes use several calibration steps usually starting with an electrically calibrated cryogenic radiometer \cite{Stock1993}. This is used to calibrate a secondary standard, which is either a photo-electric trap detector \cite{Lopez2006} or a sphere detector \cite{Eppeldauer2009}. Recently, pyroelectric detectors were tested as secondary standards and an absolute efficiency calibration with a total combined standard deviation of \textless\,0.28\,\% for wavelengths above 1\,\textmu m was achieved \cite{Alberding2022}.\\
The idea of an entirely new calibration method was presented in \cite{Vahlbruch2016}. It solely utilises squeezed states of light, the Heisenberg uncertainty principle, and relative photon loss measurements. 
The detection efficiencies of the presumably best photo diodes on the market for 1064\,nm and 1550\,nm were calibrated to $(99.5\pm0.5)$\% \cite{Vahlbruch2016} and $(98.5\pm0.7)$\% \cite{Meylahn2022}, respectively, using 15-dB and 13.5-dB squeezed vacuum states of light. 
(Both experiments maximized the efficiencies with mirrors that captured the $\approx\!0.5$\% of light reflected from the photodiodes.)
The method should be suitable for calibrating photodiodes as primary standards, but the analy\-sis of the method's precision is missing, and the current method is not without systematic errors.\\
Here, we provide the mathematical framework by establishing the mathematical relationship between the target detection efficiency and the two photoelectric measurands of the method. We derive a dedicated formula to quantify the statistical precision of the method. We demonstrate both theoretically and experimentally how all potential systematic errors can be suppressed to near-arbitrary levels. We execute the entire calibration protocol to determine the previously unknown detection and quantum efficiencies of what is arguably the most efficient photodiode currently available on the market.
We demonstrate that squeezed vacuum states of ``just'' 10-dB are suitable for calibrating the presumably most sensitive photodiodes on the market at 1550 nm to $\eta_{\rm DE}\!=\!(97.20 \pm 0.37)$\% and $\eta_{\rm QE}\!=\!(96.9 \pm 0.4)$\%. These unexpectedly low figures would make it impossible to carry out planned technology projects. (Implementing retro-reflection would increase the values by $\approx\!0.46$\%.)

\begin{figure} 
\includegraphics[width=0.94\linewidth]{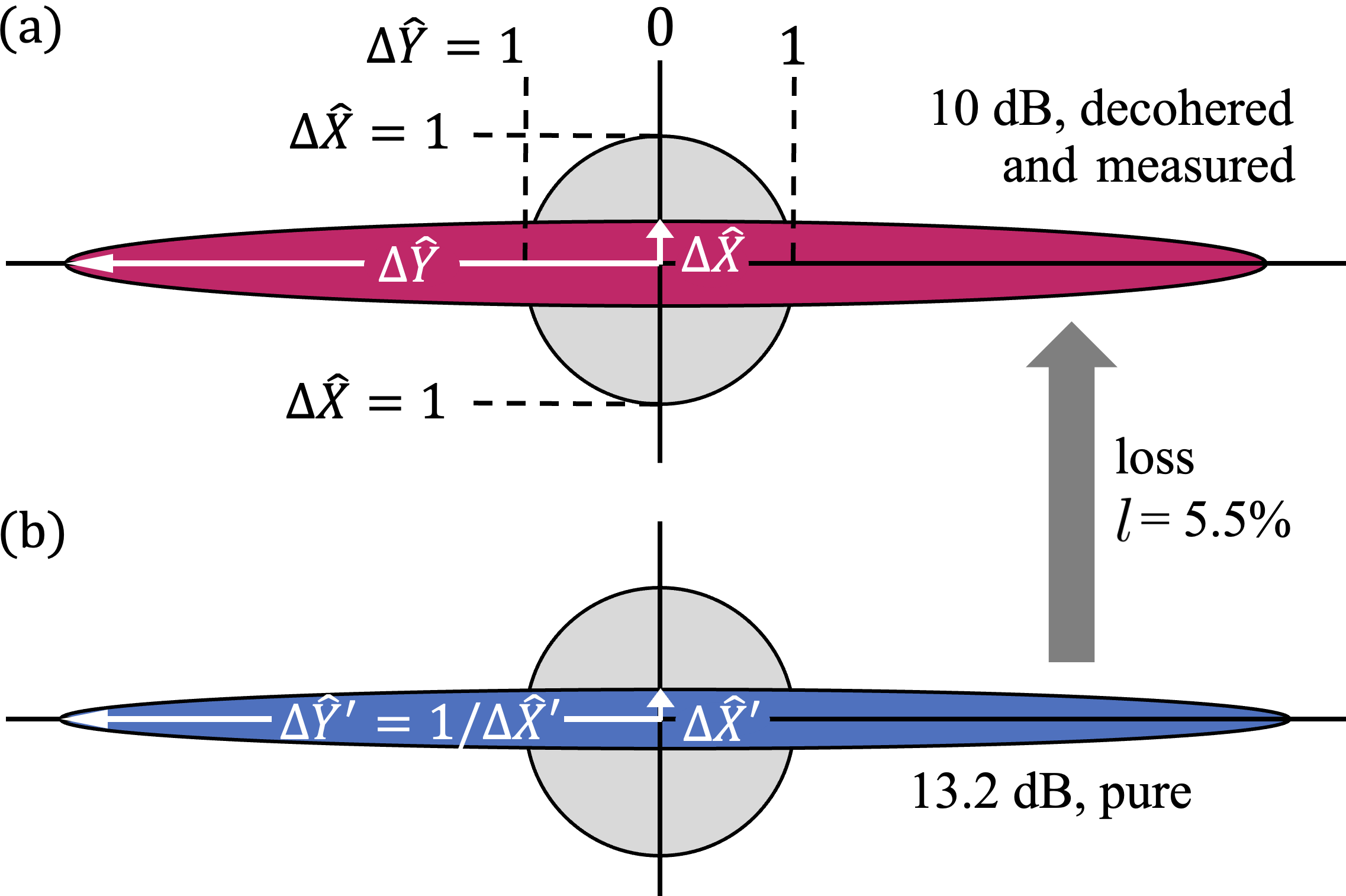}
\caption{
{\bf Decoherence due to imperfect efficiency} --- 
The ellipses represent the product of the standard deviations $\Delta \!\hat X$ and $\Delta \hat Y$ of squeezed states. The circle refers to the ground state (of the same optical mode). 
(a) Representation of the squeezed state measured in this work with imperfect total setup efficiency $\eta$.
(b) Representation of the same squeezed state as inferred for perfect efficiency (zero optical loss). 
The inference here corrects for the efficiency of $\eta = 94.5$\% (photon loss of 5.5\%) according to Eq.\,(\ref{eq:3}) yielding a higher squeeze factor and a decreased uncertainty area being at the lower bound of the Heisenberg uncertainty relation in Eq.\,(\ref{eq:2}).
}
\label{fig:1}
\end{figure}

~\\[3mm]
\noindent
{\bf 2. CONCEPT OF QRC}\\[2mm]
%
We use the following definitions of the detection efficiency $\eta_\mathrm{DE}$ and the quantum efficiency 
 $\eta_\mathrm{QE}$.  
\begin{equation}
\label{eq:1}
	\eta_{\rm DE} = \frac{U_{\rm tot} - U_{\rm dark}}{U_{\rm perf}}, \hspace{7mm}
	\eta_{\rm QE} = \frac{U_{\rm tot} - U_{\rm dark}}{U_{\rm perf} + U_{\rm dark}} \; ,
\end{equation}
where the voltages $U$ are transimpedance amplified photo currents measured on continuous-wave light. $U_{\mathrm{tot}}$ is the total measurement voltage and $U_{\mathrm{dark}}$ is the contribution without light. $U_{\mathrm{perf}}$ represents the perfect case, i.e., the unknown voltage that the measurement system would provide if every photon would be converted into exactly one photo electron, combined with zero dark current. In the limit of negligible dark noise, $\eta_\mathrm{QE}$ approaches $\eta_\mathrm{DE}$.
The challenge is to determine accurately $\eta_\mathrm{DE}$ and thus $\eta_\mathrm{QE}$, i.e., to calibrate a photodetector.

The basis of QRC is the measurement of the {two extremal variances of the quantum uncertainties of 
orthogonal electric field quadratures of a (strongly) squeezed vacuum state, $\Delta\!^2 \!\hat X$ and $\Delta\!^2 \hat Y$.}
The two obey Heisenberg's uncertainty principle \cite{Heisenberg1927, Kennard1927, Weyl1927, Robertson1929}, which, when normalised to the variance of the ground-state uncertainty, reads
\begin{equation}
\label{eq:2}
\Delta\!^2 \!\hat X \cdot \Delta\!^2 \hat Y \ge 1 \; , 
\vspace{-1mm}
\end{equation}
{where $1/(\Delta\!^2 \!\hat X) = \beta > 1$ represents the measured squeeze factor. 
(The dB-scale is given by $10\, {\rm log}_{10} \beta$). The variances are measured using 
a BHD consisting of a pair of the photodiodes to be calibrated. The result is compared with the inferred minimal uncertainty product of the {\it pure} squeezed state without decoherence.} 
Squeezed states were first described in the early 1970s \cite{Stoler1970,Lu1971,Yuen1976, Walls1983,Breitenbach1997,Schnabel2017}, first experimentally realised in the mid 1980s \cite{Slusher1985,Wu1986,Polzik1992,Vahlbruch2008}, and became a user application in 2011 when they improved the sensitivity of GEO\,600 in a joint search for gravitational waves together with Virgo \cite{LSC2011,Grote2013}.

The crucial fact is that the uncertainty product of a squeezed vacuum state initially increases successively with each lost photon, which is illustrated in Fig.\,\ref{fig:1}. Using the {setup's total photon} efficiency $\eta$, initial squeezed and anti-squeezed quadrature variances $\Delta\!^2 \!\hat X'$ and $\Delta\!^2 \hat Y' = 1/\Delta\!^2 \!\hat X'$ change  \cite{Schnabel2017} accordingly to
\begin{align}
\label{eq:3}
\!\!\Delta\!^2 \!\hat X = \eta \,\Delta\!^2 \!\hat X' + (1 \!- \eta) , \;\;\,
\Delta\!^2 \hat Y = \eta \,\Delta\!^2 \hat Y' + (1\! - \eta) \, . 
\end{align}
The variances on the left sides above can be precisely measured with a BHD {(with respect to the variance of the ground state uncertainty)} allowing for the remaining two unknown quantities $\eta$ and $\Delta\!^2 \!\hat X' = 1 / \Delta\!^2 \hat Y'$ to be calculated. Removing the latter from Eqs.\,(\ref{eq:3}) yields
\begin{align}
\label{eq:4}
\!\!\eta &= \frac{\Delta\!^2 \!\hat X+ \Delta\!^2 \hat Y - 1 - \Delta\!^2 \!\hat X \Delta\!^2 \hat Y} {\Delta\!^2 \!\hat X+ \Delta\!^2 \hat Y - 2} , \; {{\rm for} \; \Delta\!^2 \!\hat X \!,\, \Delta\!^2 \hat Y \neq 1 } \nonumber \\[1mm]
&= \frac{(1- \Delta\!^2 \!\hat X) (\Delta\!^2 \hat Y - 1)}{4 \langle \hat n_{\rm qc} \rangle} \, .
\end{align}
The two terms in the nominator represent the two required measurement results, and  $\langle \hat n_{\rm qc} \rangle = (\Delta\!^2 \!\hat X+ \Delta\!^2 \hat Y - 2)/4$ is the number of quantum correlated photons in these measurements \cite{Schnabel2017}.

The following equation splits the setup's imperfect total efficiency $\eta $ in its contributions containing the wanted detection efficiency $\eta_\mathrm{DE}$.
\begin{equation}
\label{eq:5}
\eta = \eta_\mathrm{esc}\,\eta_\mathrm{prop}\,\eta_\mathrm{mm}\,\eta_\mathrm{DE}\, .
\end{equation}
Accordingly, $\eta_\mathrm{DE}$ is found when $\eta$ and the other three efficiencies are measured. 
$\eta_\mathrm{esc}$ is the photon escape efficiency from the squeeze resonator, $\eta_\mathrm{prop}$ the propagation efficiency from the same resonator to the measurement, and $\eta_\mathrm{mm}$ the mode matching efficiency, which corresponds to the square of the interference visibility at the BHD beam splitter. 
All measurements are photoelectric in nature and take place in situ, i.e., in the same laser-optical QRC setup.
This results in high accuracy.
The squeeze resonator's photon escape efficiency $\eta_\mathrm{esc}$ could previously not be measured but just estimated \cite{Vahlbruch2016}.
The concept for measuring $\eta_\mathrm{esc}$ is one of the advancements of this work and will be discussed later. 
\begin{figure}[t!!!!!!!!!!!!!!!!!!!!!!]
\includegraphics[width=0.95\linewidth]{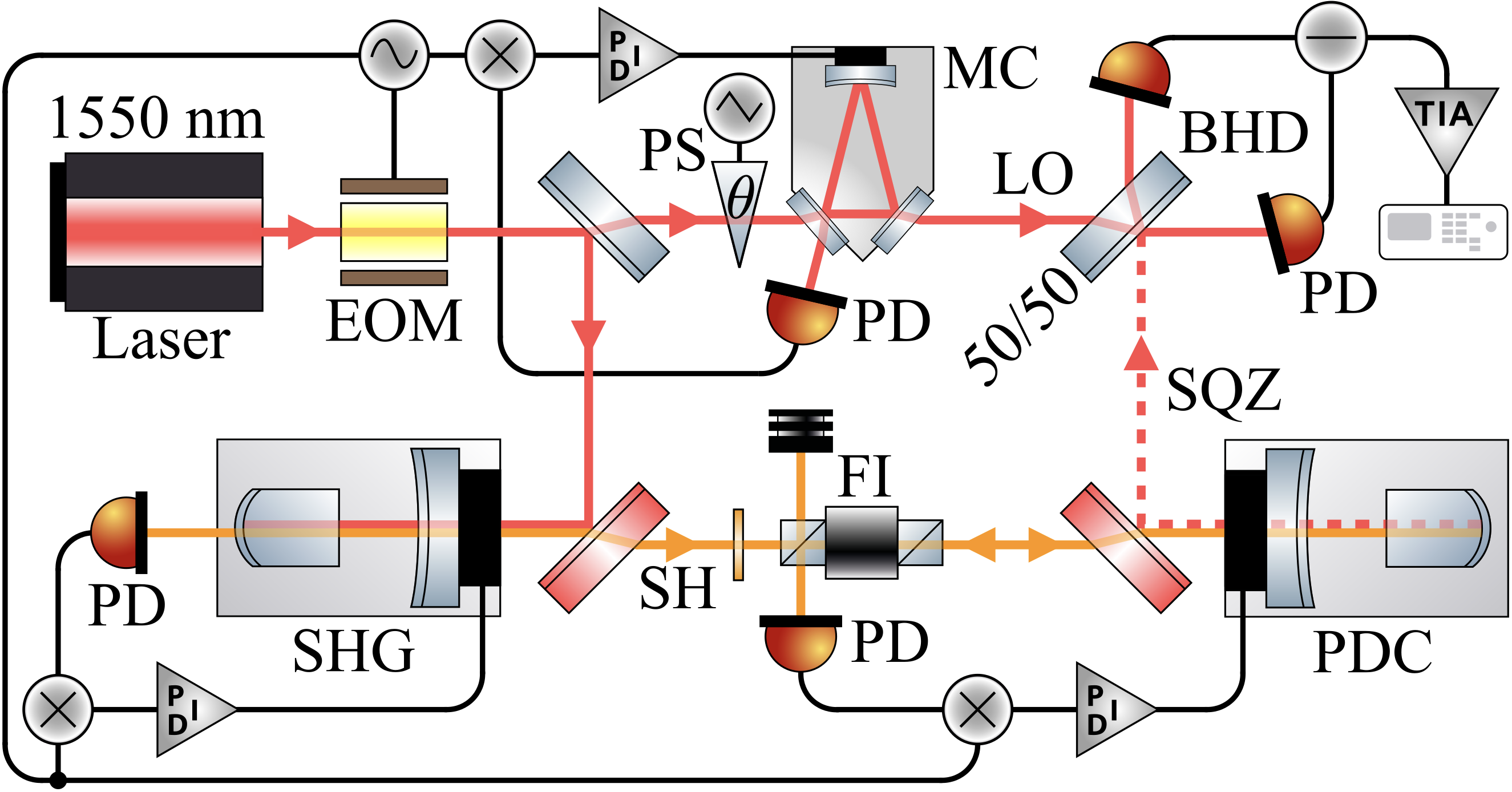}
\caption{
{\bf Schematic of the QRC setup} ---  
0.1\,W of the output power of a 1550\,nm fibre laser was converted into a 775\,nm laser beam by resonator-enhanced second harmonic generation (SHG). This light pumped our parametric down-conversion (PDC) resonator to generate squeezed vacuum states (SQZ). Two identical photodiodes (PD) to be calibrated were parts of a balanced homodyne detector (BHD) with transimpedance amplifier (TIA). The BHD's local oscillator (LO) had about 20\,mW. The BHD measured the electric field strength of the squeezed states, while the differential phase $\theta$ was continuously driven with a phase shifter (PS). The PS was placed in front of a spatial mode cleaner (MC) to avoid beam jitter on the PDs. EOM: electro-optical modulator; FI: Faraday isolator; PID: proportional-integral-derivative controller.
}
\label{fig:2}
\end{figure}

\begin{figure}
\includegraphics[width=1\linewidth]{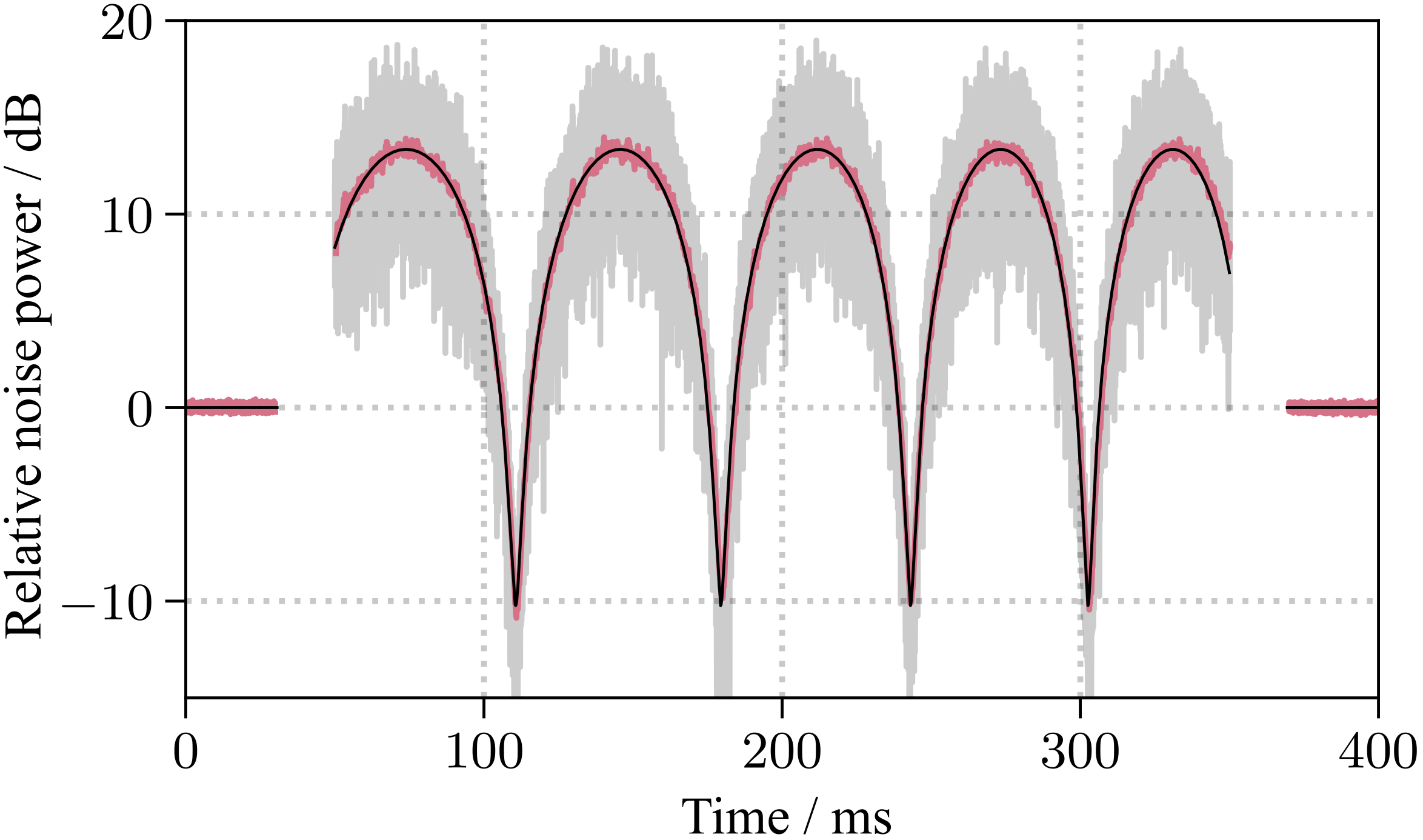}
\caption{
{\bf Measurement of the uncertainty product} --- 
Arches: Shown is a single measurement containing 32,001 data points (gray) and a model fit (black line). The readout phase $\theta$ of the BHD was swept continuously to achieve greater accuracy than previously \cite{Vahlbruch2016,Meylahn2022}. The lowest noise power corresponds to here $\Delta\!^2 \!\hat X \!\approx 0.10$, the highest to $\Delta\!^2 \hat Y \!\approx 19.7$. Noise powers are that at 5\,MHz with 300\,kHz RBW, dark noise subtracted and normalized to the noise power of the optical ground state. (The red curve shows the same data, but 500-Hz low-pass filtered on linear scale for better visualization.) 
The sweep was realized by a saw tooth voltage applied to the piezo electric element behind a steering mirror in the beam path (PS in Fig.\,\ref{fig:2}). {Horizontal lines: shot-noise.}
}
\label{fig:3}
\end{figure}

~\\[3mm]
\noindent
{\bf 3. PRECISION AND ACCURACY OF QRC}\\[2mm]
%
We present for the first time the theoretical description of the precision of the QRC method. Our result makes it clear that QRC directly exploits the quantum advantage of quantum correlations. Since $\hat X$ and $\hat Y$ do not commute, the two terms in the nominator of Eq.\,(\ref{eq:4}) must be measured one after the other. We propose here that a single measurement sequence is a zero-span measurement (single sideband frequency with finite resolution bandwidth) of 
$
\Delta\!^2 \hat X {\rm cos}^2 \theta (t) + \Delta\!^2 \hat Y {\rm sin}^2 \theta (t)$, during which the quadrature angle $\theta(t)$ is swept over several periods. Such a single measurement sequence should be mostly performed on the (strongly) squeezed vacuum state but also partly on the ordinary vacuum state providing the required reference of unity variance when the signal input port of the BHD is blocked, see Fig.\,\ref{fig:2}.
Resolution bandwidth (RBW) and sampling frequency need to be high to resolve the extremal variances of the squeezed state well. Our proposal minimizes errors due to drifts in the setup.
\\
Fig.\,\ref{fig:3} shows such a single zero-span BHD measurement with swept quadrature angle $\theta(t)$ (grey curve). 
The LO power was approximately 20\,mW and the point integration time was $\Delta T \!\approx\! 1 $\textmu s.
Each point represents an approximate measurement of $\Delta\!^2 \hat X {\rm cos}^2 \theta (t) + \Delta\!^2 \hat Y {\rm sin}^2 \theta (t)$, 
which is optically amplified by the number of photons in the LO within $\Delta T$. 
The photon number $\langle \hat n^{\Delta T}_{\rm LO}\rangle$ is that of a coherent state and equals its variance.
\\
In the following, we combine this fact with Eq.\,(\ref{eq:4}). We denote by $\langle \hat n_{\rm LO}\rangle$ the total number of photons used over many identical measurements. 
The higher this number is, the better the BHD approximation \cite{Schnabel2017} is satisfied and, consequently, the greater the precision of the variance measurement.
Fig.\,\ref{fig:3} shows one measurement, from which a value for $\eta$ is obtained. For a total measurement time of one second, this number typically lies between $10^{16}$ and $10^{18}$. Our final value is based on 100 such measurements.
The relative quantum statistical error in the calibration of $\eta$ is given by the standard error propagation formula. Assuming a precisely measured unit reference and a negligible influence from dark noise, we get
\begin{align}
\label{eq:6}
\!\!\frac{\Delta \eta}{\eta} &= \frac{\sqrt{\!\langle \hat n_{\rm LO}\rangle} \Delta\!^2\! \hat X}{\langle \hat n_{\rm LO}\rangle (1- \Delta\!^2\! \hat X)} 
+ \frac{\sqrt{\!\langle \hat n_{\rm LO}\rangle} \Delta\!^2 \hat Y}{\langle \hat n_{\rm LO}\rangle (\Delta\!^2 \hat Y - 1)}
 \nonumber \\
&= \frac{1}{\eta \sqrt{\!\langle \hat n_{\rm LO}\rangle} }
\left (1+ \frac{1-\Delta\!^2\! \hat X}{2\langle \hat n_{\rm qc} \rangle}  \right )  \, , 
\end{align}
where the last fraction tends to infinity as the squeeze factor approaches unity. 
We note that the denominator in Eq.\,(\ref{eq:4}) does not involve quantum uncertainty, is not measured, and therefore merely represents a constant factor in error propagation.
For high squeeze factors, $\Delta\!^2\! \hat X$ is negligible, which leads to
\begin{align}
\label{eq:7}
\!\!\frac{\Delta \eta}{\eta} &\lessapprox \frac{1}{\eta \sqrt{\!\langle \hat n_{\rm LO}\rangle}}
\left (1+ \frac{1}{2 \,\eta \langle \hat n'_{\rm qc} \rangle} 
\right ) \, , 
\end{align}
where $\langle \hat n'_{\rm qc} \rangle = \langle \hat n_{\rm qc} \rangle / \eta$ is the expectation value for the number of quantum correlated photons of the squeezed state before any decoherence happened \cite{Schnabel2017}. 
For the pure 13.2-dB squeezed state in Fig.\,\ref{fig:1}(b) $\langle \hat n'_{\rm qc} \rangle \approx 4.7$.\\
According to Eq.\,(\ref{eq:7}), the precision of QRC increases with the initial number of quantum correlated photons $\langle \hat n'_{\rm qc} \rangle$, the number of photons in the overlapped mode of the LO, and $\eta$. 
Our theoretical result clarifies that low-loss optics should be used in QRC and that higher precision is achieved for photo diodes with higher $\eta_{\rm DE}$.
(A limit to $\langle \hat n'_{\rm qc} \rangle$ is set by the fact that the decoherence due to phase fluctuations must remain negligible, see `methods').  

~\\[3mm]
\noindent
{\bf 4. QRC WITH 10-dB SQUEEZED LIGHT}\\[2mm]
%
We applied our QRC to a pair of 1550-nm photodiodes from {\it Laser Components} (`HQE', with removed protection windows), which were optimized for s-polarized continuous-wave laser light at 1550\,nm under 20$^\circ$\,angle of incidence. Their semiconductor surfaces were free of dust particles. The two photodiodes formed a balanced homodyne detector for measuring $\Delta\!^2 \!\hat X$ and $\Delta\!^2 \hat Y$ of a squeezed state for the determination of $\eta$ according to Eq.\,(\ref{eq:4}). These measurements are not performed at direct current, but at a sideband frequency (within the squeezing bandwidth) for which the photodiodes are designated. Here, we chose 5\,MHz. We used resonator-enhanced degenerate parametric down-conversion (PDC) in a second-order non-linear crystal for squeezed light generation \cite{Wu1986,Polzik1992,Vahlbruch2008,Vahlbruch2016,Schoenbeck2018}, see Fig.\,\ref{fig:2}.
The crystals in the SHG and squeeze resonators were periodically poled potassium titanyl phosphate (PPKTP). The crystal geometry was plane-convex and formed a half-monolithic resonator together with a coupling mirror. The crystal was coated to be highly reflective on the convex side and anti-reflective on the other for 1550\,nm and 775\,nm. The coupling mirror had a design reflectivity of $\approx\! 83$\,\% for 1550\,nm and $\approx\! 97.5$\,\% for 775\,nm light. The temperature profile of the crystal  \cite{Lastzka2007,Schoenbeck2018phd, Hagemann2024} was stabilized to provide double-resonance as well as good phase matching. The length of the squeeze resonator was stabilized using the Pound-Drever-Hall (PDH) technique \cite{Drever1983}. The squeezed vacuum states at 1550\,nm (without any carrier light) `escaped' the squeeze resonator through the coupling mirror, were separated from the 775\,nm light via a dichroic mirror, and superimposed with the local oscillator on the BHD's beam splitter. We directly measured a 10-dB squeezed photon shot noise. 
The anti-squeezing was close to 13\,dB. Increasing the pump power increased the squeeze factor just marginally, while the anti-squeezing increased significantly. We used this measurement to estimate the root mean square (rms) phase noise \cite{Aoki2006} to $\Delta \theta\!<\!(2 \!\pm\! 1)$\,mrad, which was negligible for the QRC result reported here, see `methods'.\\ 
%
Fig.\,\ref{fig:3} shows the {rms-}measurement of $\Delta\!^2 \!\hat X$ (minima) and $\Delta\!^2 \hat Y$ (maxima) for determining the $\eta$. {\it Sweeping} the BHD read-out phase $\theta$ offers a few advantages over the alternative of {\it locking} the read-out phase to {$\Delta\!^2 \!\hat X$ and $\Delta\!^2 \hat Y$} as used before \cite{Vahlbruch2016,Meylahn2022}. First of all, it reduces the complexity of the setup by omitting the stabilization of one degree of freedom. Additionally, slow drifts{ in the local oscillator and pump power can be neglected. Scanning the read-out phase also guarantees that there is no systematic error in the locking point and ensures that each quadrature is optimally measured at some point. The traces in Fig.\,\ref{fig:3} already include dark noise subtraction and normalisation to the ground state variance. The variance of the ground state was measured with the signal input of the BHD blocked. The variance of the dark noise was {20\,dB below shot noise and} measured with all light fields blocked.\\
The traces in Fig.\,\ref{fig:3} were fitted with the model 
\begin{eqnarray}
\hspace*{-8mm}
\label{eq:8}
P({\theta}(t)) =\,&& \Delta\!^2\!X \cos^2 \!\theta(t) \!+\! \Delta\!^2 Y \sin^2 \!\theta(t) \\ 
{\theta}(t) =\,&& {\theta_0} + {\theta_1} t + {\theta_2} t^2 \, ,
\label{eq:9}
\end{eqnarray} 
where the phase evolution of the figure's x-axis according to Eq.\,(\ref{eq:9}) was given by the piezo expansion.
Fig.\,\ref{fig:3} shows one of altogether 100 measurements. 
Every measurement provided an efficiency value $\eta$ using Eq.\,(\ref{eq:4}).
The statistics provided $\eta = (94.48\,\pm\,0.22)$\%.\\
For the $20^\circ$-angle of incidence used, $(0.46\pm0.06)$\% of the light was reflected from the photo diodes. 
A retro-reflector would have increased the values of $\eta$, $\eta_\mathrm{DE}$, and $\eta_\mathrm{QE}$ accordingly. Most likely, our error specification would have {\it decreased} minimally, see Eq.\,(\ref{eq:6}).

The {\it escape efficiency} $\eta_{\rm esc}$ of the squeeze resonator could previously only be determined with low precision and low accuracy due to potential systematic errors \cite{Vahlbruch2016, Meylahn2022}. Here, we apply an in-situ measurement with high precision and accuracy. In particular, potential surface inhomogeneities and small dust particles in the squeeze resonator, which make the value of $\eta_{\rm esc}$ dependent on its alignment, are taken into account.\\ 
The escape efficiency of the quantum correlated photons from the squeeze resonator is given by 
\begin{eqnarray}
\eta_{\rm esc} = \frac{1-r^2}{1-r^2 + \ell_\mathrm{rt}^2}  \, ,
\label{eq:10}
\end{eqnarray}
where $r^2$ is the reflectivity of the ramped coupling mirror in Fig.\,\ref{fig:4} and $\ell_\mathrm{rt}^2$ is the cavity's relative photon loss per roundtrip, with contributions $\ell^2_\mathrm{scat}$, $\ell^2_\mathrm{abs}$, $\ell^2_\mathrm{trans}$, and $\ell^2_\mathrm{refl}$.
\begin{figure}
\hspace*{1mm}
\includegraphics[width=0.95\linewidth]{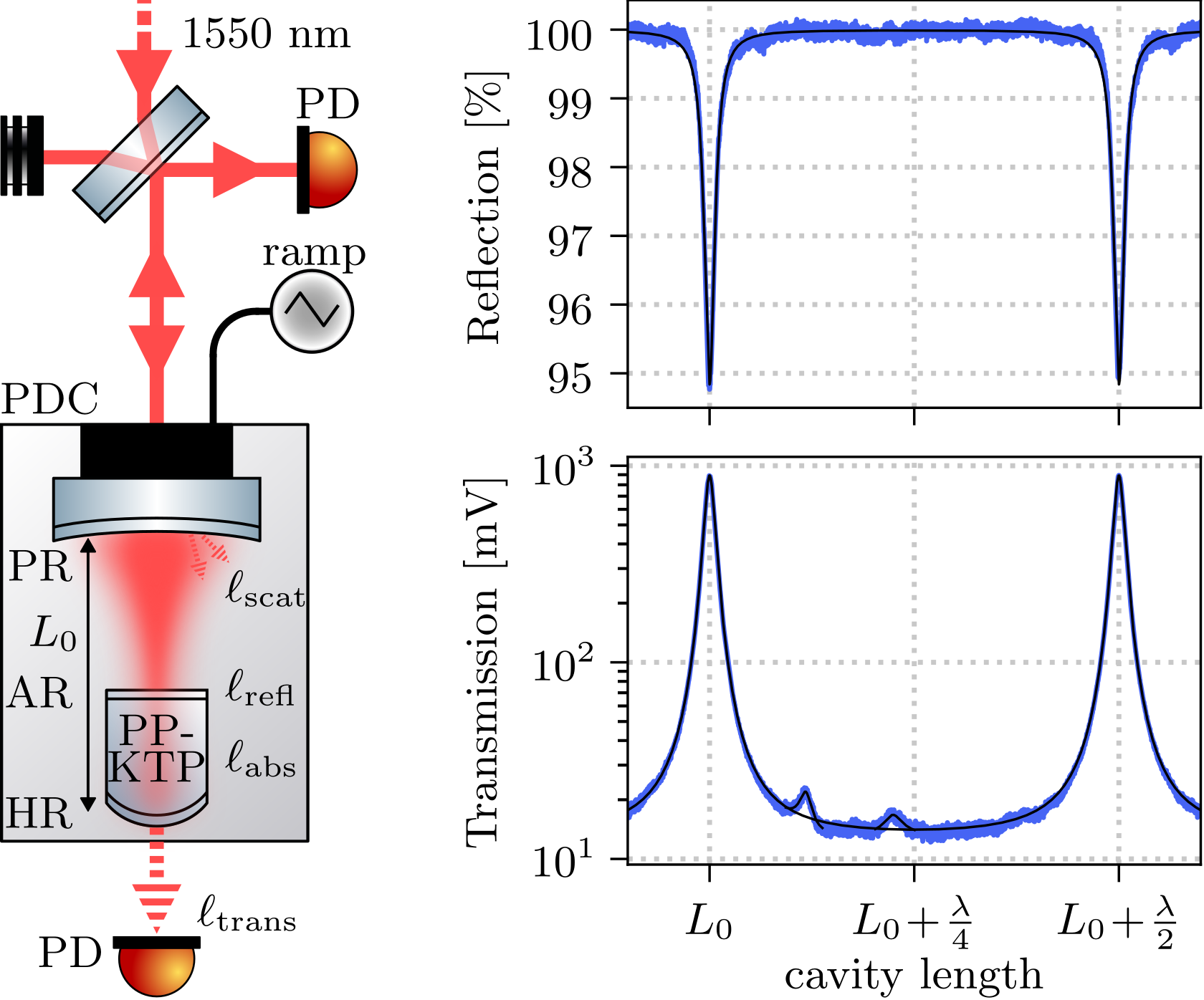}
\caption{
{\bf Determining the escape efficiency} ---
Left: squeeze resonator whose escape efficiency $\eta_{\rm esc}$ needs to be determined in a separate measurement. 
$\eta_{\rm esc}$ is imperfect due to intra-cavity photon scattering ($\ell^2_\mathrm{scat}$) and absorption ($\ell^2_\mathrm{abs}$) at all three intra-cavity surfaces and during crystal transmission. Further photon losses are the residual transmission ($\ell^2_\mathrm{trans}$) and reflection ($\ell^2_\mathrm{refl}$) of the convex and plane crystal surfaces, respectively. Exactly the same loss sources also reduce the reflectivity of the resonator according to Eq.\,(\ref{eq:11}), which is accessible through an in-situ measurement of mode-matched, retro-reflected light.
{AR, PR, HR: anti-, partial, and high-reflectivity coating.}
Top right: Measurement of the retro-reflected light while the resonator's length is scanned. Dips occur on resonance when the light couples and experiences the intra-cavity loss (to be up-scaled by the imperfect mode matching). The black line in the measurement graph is the theoretical model with fitted coupling mirror reflectivity and roundtrip loss.
Bottom right: Simultaneous measurement of transmitted light provides the imperfect mode matching. 
}
\label{fig:4}
\end{figure}
Our `in-situ'-method measures $r^2$ and $\ell_\mathrm{rt}^2$ on the finally aligned squeeze resonator, thus including potential local scattering and absorbing defects. We first mode-matched a dim auxiliary 1550-nm field into the squeeze resonator from its over-coupled side. The retro-reflected beam was monitored by a photodiode while the resonator length was scanned with a piezoelectric element and external 775-nm light blocked. A second PD (which was insensitive to 1550\,nm light) simultaneously confirmed that no second harmonic light was produced. To achieve this, the temperature of the crystal was set to a conversion minimum under strong pumping. 
The light power reflected from the scanned resonator reads 
\begin{eqnarray}
I(\varphi(t)) = \left|\frac{r - \sqrt{1 - \ell_\mathrm{rt}^2} e^{-2i\varphi(t)}}{1 - r  \sqrt{1 - \ell_\mathrm{rt}^2} e^{-2i\varphi(t)}}\right|^2 \!I_0\, ,
\label{eq:11}
\end{eqnarray}
where {$\varphi(t)$ is the detuning of the squeeze resonator and} $I_0$ is the power of the incident beam. 
Fig.\,\ref{fig:4} shows our setup for determining $\eta_{\rm esc}$\,(left), a measurement graph of the reflected light power to which we fitted Eq.\,(\ref{eq:11}) (top right), and the simultaneously measured transmitted power (bottom right).
The latter provided a cavity mode matching value of $(98.58 \pm 0.23)$\% (main peak divided by sum of all peaks), which was required to scale up the loss dip in the top right graph.
The fit of Eq.\,(\ref{eq:11}) to the up-scaled top right graph required a simultaneous fit of the scanning speed, which we approximated by $\varphi(t) = \varphi_0 + \varphi_1 t + \varphi_2 t^2$.
Our in-situ measurement yielded $r^2 = (82.79 \pm 0.35)$\% and $\ell^2_\mathrm{rt} = (0.247 \pm 0.007)$\% 
and thus $\eta_{\rm esc} = (98.583 \pm 0.015)$\% according to Eq.\,(\ref{eq:10}).\\[-2mm]

{\it Measurement of the propagation efficiency} --
We blocked the second harmonic pump beam and coupled an auxiliary 1550-nm field through the PDC cavity from the under-coupled side. The transmitted light on cavity resonance was in the same mode as the squeezed vacuum and experienced the same optical losses along the down stream path. Measuring its power behind the coupling mirror $P_\mathrm{cm}$ and in front of each homodyne photodiode $P_\mathrm{PD1, PD2}$ allowed for determining the propagation efficiency $\eta_\mathrm{prop} = (P_\mathrm{PD1} + P_\mathrm{PD2}) / P_\mathrm{cm}$. The optical loss of the coupling mirror itself was dominated by its anti-reflective coating with AR = 500\,ppm\,$\pm$\,500\,ppm. Our measurement yielded $\eta_\mathrm{prop} = (99.49\,\pm\,0.25)$\%.\\[-2mm]

{\it Measurement of the mode matching efficiency} --
We used the same auxiliary beam as before, precisely adjusted its power to match that of the LO, and measured the interference contrast at the beam splitter of the BHD.
The differential phase $\theta$ was swept linearly by a piezo-actuated steering mirror and the resulting fringes were recorded using a single photodiode of the BHD. The fringe visibility was derived according to $v \!\!\;=\!\!\; (U_{\rm max} \!\!\;-\!\!\; U_{\rm min})/(U_{\rm max} \!\!\;+\!\!\; U_{\rm min})$, where $U_{\rm max,min}$ were the photoelectric voltages after subtraction of the voltage without any light. Our measurement yielded $\eta_\mathrm{mm} = v^2 = (99.11 \,\pm\, 0.17)\%$.\\[-2mm]

\begin{figure}[h!!!!!!!!!!!!!!!!!]
\includegraphics[width=1\linewidth]{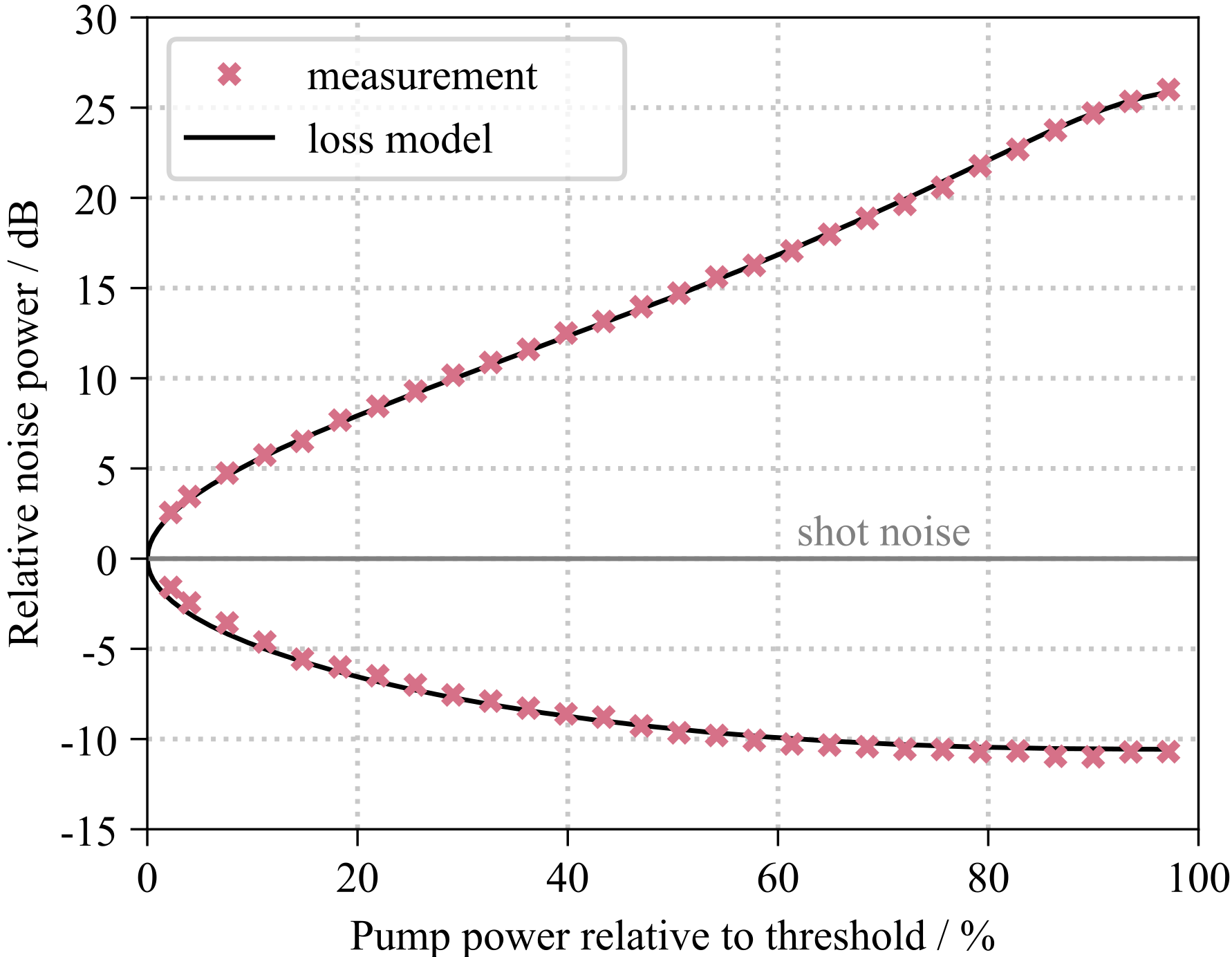}
\caption{
{{\bf Variation of pump power} --- The measured squeeze and anti-squeeze values are in very good agreement with a model that assumes a constant loss as the sole decoherence mechanism. Vertical error bars are smaller than the crosses.}
}
\label{fig:5}
\end{figure}
The QRC decoherence model, see Fig.\,\ref{fig:1}, assumes constant optical loss and no additional decoherence mechanisms. It must be verified experimentally that this assumption holds.
Fig.\,\ref{fig:5} shows our measured squeeze and anti-squeeze values versus pump power, alongside the theoretical model that assumes a constant optical loss and no additional decoherence effects. The agreement is very good. 
We conclude that, even within an extended measurement range of our experiment, neither phase noise, nor thermal deformations, nor any other effects dependent on the pump power occur.

~\\
\noindent
{\bf 5. CALIBRATION RESULT}\\[2mm]
%
The two photodiodes that were calibrated in this work have an average detection efficiency of 
\begin{align}
\eta_\mathrm{DE} ({\rm 1550\,nm}) =\,& \frac{(94.48\!\!\;\pm\!\!\;0.22)\% \,/ (98.583\!\!\;\pm\!\!\;0.015)\%}{\!\!(99.49\!\!\;\pm\!\!\;0.25)\% \times (99.11\!\!\;\pm\!\!\;0.17)\%} \nonumber \\
=\,&\,(97.20\!\!\;\pm\!\!\;0.37)\%. \nonumber
\label{eq:10}
\end{align}
Our error bar corresponds to the total combined standard deviation. It is dominated by statistical errors in the individual measurements and can be further reduced by increasing the measurement time, using lower-loss optics and also fewer optics for beam guidance and focusing on the photodiodes. Already with this experiment we almost achieve the combined standard deviation reported in Ref.\cite{Alberding2022}.
Unlike \cite{Alberding2022}, we exclusively use the photoelectric effect and not the pyroelectric effect, for which light must heat a material through absorption.\\ 
The quantum efficiency is derived from the detection efficiency, taking into account the measured dark noise at 5\,MHz, see Eq.\,(\ref{eq:1}). For the local oscillator power of 10\,mW used, we obtain the only slightly lower value of $\eta_\mathrm{QE}{\rm (10\,mW, 5\,MHz)} = {(96.9\!\!\;\pm\!\!\;0.4)}\%$. We note that in our previous work \cite{Vahlbruch2016}, we only specified the value for detection efficiency according to Eq.\,(\ref{eq:1}).

~\\
\noindent
{\bf 6. SUMMARY AND CONCLUSIONS}\\[2mm]
%
Our work shows that QRC is based on quantum-correlated photons and that its precision increases with their number. This clarifies that QRC directly exploits the advantage of quantum correlations. We exclude the potential sources of systematic error, and, furthermore, present a calibration example that is of great interest to researchers and developers of quantum technologies.  We calibrate the detection efficiency of just $(97.20\!\!\;\pm\!\!\; 0.37)\%$ for the presumably best photodiodes for 1550\,nm on the market (Laser Components). The value is clearly too low for optical quantum computers and for the current design of the Einstein Telescope (ET). Furthermore, it resolves the long-standing discrepancy in the maximum achievable squeeze factors at 1064\,nm \cite{Vahlbruch2016} and 1550\,nm \cite{Schoenbeck2018,Meylahn2022}. The pair of photodiodes calibrated here is considered a primary standard, allowing other photodiodes to be directly compared and calibrated against it.
\\
Based on these breakthroughs, we consider it likely that our calibration method will become established, serving as the second example of a true user application of quantum correlations, following squeezed light for improving gravitational wave detection. Squeeze lasers of the required quality are commercially available \cite{NoisyLabs}. Our analysis shows that the precision of QRC also increases with the detection efficiency of the photodiode, meaning it could become an important benchmarking tool for the realisation of fault-tolerant optical CV quantum computers \cite{Larsen2019,Fukui2022,Konno2024,Larsen2025}.\\

\noindent 
{\bf Acknowledgments}\\[1mm]
This work was supported and partly financed (L.A.) by the DFG under Germany's Excellence Strategy EXC 2121 ``Quantum Universe'' -- 390833306.\\

\noindent 
{\bf Author Contributions}\\[1mm]
L.A. performed the final experiments and the data analysis. J.-M.M. performed initial $\eta_{\rm esc}$-measurements. R.S. invented QRC and proposed the $\eta_{\rm esc}$-measurement approach, developed the theoretical QRC description, and prepared the manuscript with the help of L.A.\\

\noindent 
{\bf Disclosures}\\[1mm]
The authors declare no conflicts of interest.\\

\noindent 
{\bf Data availability}\\[1mm]
The experimental datasets will be available at ... .\\

\noindent 
{\bf Supplemental document}\\[1mm]
See Supplement 1 for supporting content.


\newpage

\noindent
{\Large \bf{Supplement}}
~\\

\noindent
{\bf A. BALANCED HOMODYNE DETECTION}\\[2mm]
%
The most important test for ruling out systematic errors caused by the BHD is to verify its proportionality. The measured spectral noise power must be proportional to the LO power across the entire target measurement range when the signal input is blocked, i.e., when the vacuum state is measured (see for instance Fig.\,2 in Ref.\,\cite{Vahlbruch2007}). If the noise power increases more rapidly, it is not the noise power of the ground state that is being measured, but that of other noise sources, e.g., scattered photons. If it increases more slowly, this is an indication of saturation of the amplifier electronics.\\
Regardless of the BHD, a systematic error can arise if the phase $\theta$ of the squeezed light fluctuates during the measurement \cite{Franzen2006}. This type of decoherence also leads to a reduction in the squeeze factor, even if all photons are detected. By operating the squeeze resonator close to its oscillation threshold, the phase noise becomes visible and can be quantified (see for instance Fig.\,4 in Ref.\,\cite{Mehmet2011}).
Ideally, QRC uses a range of $\langle \hat n_{\rm qc} \rangle$ in which phase fluctuations are demonstrably irrelevant.\\

\noindent
{\bf B. CALIBRATING AN UNBALANCED PAIR OF PHOTO DIODES}\\[2mm]
%
Ideal balanced homodyne detection (BHD) is the noise-less optically amplified measurement of the field quadrature  $\hat X_\theta = \hat X {\rm cos} \theta + \hat Y {\rm sin} \theta $, where $\theta$ is the adjustable relative phase of the local oscillator, see Figs.\,2 and 3. If the detection efficiency of the two photodiodes is imperfect ($\eta_{\rm DE} < 1$) but identical, the usual BHD approximation \cite{Schnabel2017} yields
\begin{equation*}
\Delta^{\!2}(\hat o_1^\dagger \hat o_1 - \hat o_2^\dagger \hat o_2) 
  \approx \eta_{\rm DE} |\alpha|^2 \left(\eta_{\rm DE} \Delta^{\!2}\! \hat X_\theta + 1-\eta_{\rm DE}\right) \, ,
\end{equation*}
where $\hat o_{\rm i}^\dagger$ and $\hat o_{\rm i}$ are the creation and annihilation operators of the mode measured by photo diode `i' after loss, and $|\alpha|^2$ is the average photon number of the local oscillator mode approximated by a coherent state.\\
If the quantum efficiencies of the two photodiodes differ ($\eta_{\rm DE1} \neq \eta_{\rm DE2}$), the experiment must  compensate for that by a suitably unbalanced beam splitter in order to maintain balanced operation for noiseless optical amplification, i.e.
\begin{equation*}
0 = \eta_{\rm DE1} r^2 \! - \eta_{\rm DE2} t^2 \, \approx \, \langle \hat o_1^\dagger \hat o_1 \rangle - \langle \hat o_2^\dagger \hat o_2 \rangle  \, .
\end{equation*}
~
\\[3mm]
In our experiment, we rotated the BHD's beam splitter until balance was achieved, without obtaining any information about a potential difference in detection efficiencies. 
The above equation combined with that of the loss-less beam splitter ($r^2 + t^2 = 1$) provides 
\begin{align*}
\Delta^{\!2}(\hat o_1^\dagger \hat o_1 &-  \hat o_2^\dagger \hat o_2) \approx \frac{\eta_{\rm DE1} + \eta_{\rm DE2}}{2} |\alpha|^2 \\
  & \times 
   \left( \frac{\eta_{\rm DE1} + \eta_{\rm DE2}}{2} \Delta^{\!2} X_\theta + 1 
         -  \frac{\eta_{\rm DE1} + \eta_{\rm DE2}}{2} \right) \, .
\end{align*}
This shows that the QRC method provides the arithmetic mean of the detection efficiencies of the two photo diodes under consideration. A simple measurement of the difference in the optical DC power levels at the photodiodes would determine the difference in $\eta_{\rm DE1}$ and $\eta_{\rm DE2}$ on absolute scale. 
Our result that QRC calibrates the average value of two photodiodes is the desired one. Calibration therefore yields a value that is statistically representative of a pair (but which may also be representative of a production batch of many identical photodiodes). In any case, it is usually the detection efficiency of the entire BHD system that is of interest.

\end{document}